\documentclass[runningheads]{llncs}
\usepackage{graphicx}
\usepackage{enumitem}
\usepackage{mathtools}
\usepackage{bussproofs}
\usepackage{stackengine}
\usepackage{amsfonts}
\usepackage{listings}
\usepackage{color}
\usepackage{js-and-yaml}

\makeatletter
\renewcommand{\paragraph}{%
  \@startsection{paragraph}{4}%
  {\z@}{0.7ex \@plus 1ex \@minus .2ex}{-1em}%
  {\normalfont\normalsize\itshape}%
}
\makeatother

\makeatletter
\def\@seccntformat#1{\@ifundefined{#1@cntformat}%
   {\csname the#1\endcsname\quad}  
   {\csname #1@cntformat\endcsname}
}
\let\oldappendix\appendix 
\renewcommand\appendix{%
    \oldappendix
    \newcommand{\section@cntformat}{\appendixname~\thesection\quad}
}
\makeatother

\usepackage{hyperref}

 \newcommand{\myfigureshrinker}{\vspace{-.8cm}}
 \newcommand{\bottomfigureshrinker}{\vspace{-.4cm}}

\definecolor{teal}{rgb}{0.1,0.6,0.4}

\begin{document}
\title{Formalizing Event-Driven Behavior of Serverless Applications}
%
%
\author{Matthew Obetz \and
Stacy Patterson \and
Ana Milanova}
\authorrunning{M. Obetz et al.}
%
\institute{Rensselaer Polytechnic Institute, Troy NY 12180, USA 
\email{\{obetzm,pattes3,milana2\}@rpi.edu}}
\maketitle              
\begin{abstract}
We present new operational semantics for serverless computing that model the event-driven relationships between serverless functions, as well as their interaction with platforms services such as databases and object stores. These semantics precisely encapsulate how control transfers between functions, both directly and through reads and writes to platform services.
We use these semantics to define the notion of the service call graph for serverless applications that captures 
 program flows through functions and services.
 Finally, we construct service call graphs for eight serverless JavaScript  applications, using a prototype of our call graph construction algorithm, and  we evaluate their accuracy.

\end{abstract}


\vspace{-.8cm}
\section{Introduction}
\vspace{-.2cm}

Serverless computing is a programming model where code executes on-demand in a shared network of pre-configured computing resources~\cite{lynn2017preliminary}.  By pooling resources and managing the execution platform, serverless platform providers are able to offer highly elastic scaling by balancing workloads across multiple physical servers~\cite{perez2018serverless}. This distribution across physical servers is made possible through the use of  \textit{containers} that bundle serverless code with a virtual execution environment. Serverless providers can dynamically scale the number of instances of a serverless application by creating multiple containers.
To facilitate this approach, serverless application code is organized into stateless \textit{serverless functions} that execute in response to events. As a consequence, serverless computing heavily favors microservice architectures; serverless functions pass messages and subscribe to notifications from other platform services to complete tasks cooperatively~\cite{spillner2017practical}. The advantages of this platform have led to the rapid adoption of commercial serverless platforms by developers in industry~\cite{baldini2017serverless}.

However, the serverless model also presents new challenges. In particular, previous research has cited a lack of tooling for development and debugging of serverless applications~\cite{yan2016building}. Without access to these tools, developers may struggle to trace executions, measure performance, and verify the security of programs they write. While significant progress has been made toward answering these questions for traditional programs using program analysis, this analysis has not yet been significantly extended to work in the serverless domain.
Recent surveys of the state of serverless computing have suggested that static analysis can help address these challenges~\cite{berkeley2019cloud}.

Existing abstractions for serverless computing emphasize unique features of the environment where serverless functions are executed~\cite{Jangda2019Formal,Gabbrielli2019nomore}. However, these abstractions do not consider effects of transmitting data to other services and functions.
Data transmitted in this fashion is commonly replicated to new executions of serverless functions that spawn in response to a change in state on their associated service. Without \emph{operational semantics} that capture this behavior, program analysis cannot construct a precise call graph and cannot soundly reason about dataflow between parts of a serverless application. The lack of semantics to describe these \textit{event triggers} also serves as a barrier to more advanced reasoning about data privacy, application correctness, and resource usage. 

To address this gap, we propose new operational semantics for event-driven serverless computation. These semantics describe how writes and reads to platform services create inter-function control transfer in serverless applications. Our semantics formalize the most common platform services including object stores, databases, notifications, queues and stateless services. We then define a new approach to call graph construction for serverless applications that uses these semantics to augment call graphs with information about relationships between serverless functions and platform services. We introduce the notion of the \emph{service call graph}, which extends the classical call graph to include new nodes. These new nodes represent the platform services written to or read by application code to produce control flow that spans multiple disconnected parts of a program. By tracing control flow through reads and writes to services, individual serverless functions become a single unified application with additional context describing what data may flow to later functions in a call chain.

We make the following contributions:
\begin{itemize}[noitemsep,nolistsep,leftmargin=*]
	\item We formulate new operational semantics for the execution of serverless programs. These semantics precisely model interactions with platform services, including event triggers that causes additional functions to execute. 
	\item  We extend the traditional notion of a call graph with new types of nodes and edges that represent event-driven behavior on serverless platforms. These new nodes and edges capture the inter-function control and state transfer represented in our operational semantics.
	\item  We design and implement an algorithm for constructing call graphs of serverless programs. We evaluate the accuracy of our approach by presenting metrics on the call graphs produced by a prototype implementation of our algorithm against  serverless programs collected from GitHub.
	We focus on applications written in Javascript for the AWS Lambda platform~\cite{awslambda}; we choose Javascript as it is the most common language for AWS Lambda programs.  
\end{itemize}

\vspace{-.5cm}
\subsubsection{Related Work.}

The semantics we define for the lifecycle of a single serverless function are closely related to those used in a recent formalization of serverless computing~\cite{Jangda2019Formal}. That work focused on modeling low-level behavior of serverless systems. Such models are useful for capturing behavior such as program non-determinism that can arise from reading state from previous executions of serverless functions. Our semantics start from this model to describe initiating requests, language-agnostic computation steps, and generated responses. However, the semantics defined in \cite{Jangda2019Formal} do not capture inter-function communication and program flows that span multiple serverless functions. Specifically, these semantics limit data persistence to a locking transactional key-value store. Our semantics introduce several new state domains that model the behavior of these services. More importantly, the previous semantics also lack a conceptualization of serverless events, which initiate execution of a serverless function when state is manipulated on a data storage service. We model these interactions by extending the semantics with a new collection of event semantics that capture state transfer between serverless components.

Dynamic analysis has also been previously explored as a tool for reasoning about the behavior of serverless applications. Specifically, this line of research has developed systems to visualize program structure~\cite{lin2018tracing,lin2018tracking}, track the flow of sensitive information~\cite{alpernas2018secure}, and measure resource costs~\cite{wang2018curtain}. These systems instrument new tools that modify or extend serverless platforms with runtime logging and label checking that enable them to make partial judgments about security given their partial view of current system state. By contrast, we are interested in formalizing serverless behavior so that such analyses may be performed statically without requiring application deployment or execution.

The service call graph shares some features of message flow graphs for distributed event-based systems that communicate through publish-subscribe middleware~\cite{Garcia2013identifying}. 
The publish-subscribe model is related to the serverless notification systems, however, 
retrieval of data from databases and object stores cannot be succinctly captured in publish-subscribe semantics. Our work considers not only notification-based communication, but also messages that pass through other channels available to serverless applications.

In preliminary work~\cite{hotcloud}, we introduced the notion of a service call graph for serverless applications. In this paper, we formalize the call graph definition in terms of our new operational semantics. Further, we design and implement a call graph construction algorithm and present experimental results on eight real-world serverless applications.   

\vspace{-.5cm}
\subsubsection{Outline.} 
The remainder of this paper is organized into the following sections. Section~\ref{sec:model} defines the serverless computing model, then Section~\ref{sec:semantics} maps this model onto a set of operational semantics that formalize serverless computation.
Section~\ref{sec:callgraph} presents our a serverless call graph construction. We evaluate the accuracy of call graph construction in Section~\ref{sec:implementation}, and present conclude in Section~\ref{sec:conclusion}.


\section{The Serverless Model}
\label{sec:model}

Serverless computing is a new programming model that allows developers to execute modules of code in a distributed setting without specifying the physical servers where this code will run. 
 The most common implementation of serverless computing is the Function-as-a-Service (FaaS) model. 
In this model, an application is decomposed into a collection of \textit{serverless functions}. 

Most platforms provision resources for serverless functions by deploying virtual containers to execute function code~\cite{Dua2014Virtualization}. These containers provide a minimal operating system, including a language interpreter specified by the function runtime. Since these containers can be quickly created and destroyed on-demand, serverless platforms are able to reuse the same physical hardware to service multiple functions, even from different users. Additionally, many serverless platforms are optimized to allow the container of a frequently-called function to remain provisioned. A \textit{warm start} occurs when a function is invoked from a reused container. Warm starts eliminate the latency associated with waiting for new containers to initialize during a \textit{cold start}, but create a risk of stale memory allocations from previous invocations of the function affecting program behavior.

Serverless functions interact with one another either via direct invocation or via \textit{services} that expose interfaces for data storage and messaging. 
Below, we present the set of common services provided by serverless platforms, followed by a description of the methods by which larger serverless applications can be constructed from functions and services.

\vspace{-.2cm}
\subsection{Platform Services}

Services and functions interact in two main ways: 1) functions write data to services through the use of platform-provided libraries, and 2) functions receive data from services, either as part of an explicit read using those same libraries, or as inputs assigned to the parameters of a function when it is initially invoked. We define five broad categories of services.
	
\paragraph{Object stores.}  Object stores persist unstructured data in buckets. Each item uploaded to a bucket is identified by a unique key. This key can be used to retrieve the item for reading.
Object stores are commonly used as a replacement for a filesystem in serverless applications. Examples of object store services include  Amazon Simple Storage Service~\cite{s3}, Google Cloud Storage~\cite{cloudstorage}, and Azure Blob Storage~\cite{azureblob}. 

\paragraph{Databases.} Databases store semi-structured data in one or more tables. Unlike object stores, databases provide advanced APIs for retrieving data based on queries. This category includes both relational databases such as Amazon Aurora~\cite{aurora}
and Azure Cosmos DB~\cite{cosmos}, as well as column-store NoSQL databases such as DynamoDB~\cite{dynamo}.

\paragraph{Notifications.} Notification services expose a collection of named \textit{topics}, which may be organized into a hierarchy for granular filtering. When a serverless function \textit{publishes} data under a topic, all functions that \textit{subscribe} to that topic or a parent topic are invoked and receive a copy of the data. Commercially available notification services include Amazon Simple Notification Service~\cite{sns}, Google Cloud Pub/Sub~\cite{cloudpubsub}, and Firebase Cloud Messaging~\cite{firebase}. 

\paragraph{Queues.} Queues allow for intermediate storage of data that requires further processing. Serverless functions can be configured to execute when a queue receives new items. 
Queues either invoke serverless functions immediately when data is added to the queue, or batch several queued items in an array to be processed by a single invocation of a serverless function.  Example serverless queues include the Amazon Simple Queue Service~\cite{sqs} and Amazon Kinesis Streams~\cite{kinesis}. 

\paragraph{Stateless Services.} Stateless services perform data processing on-demand for serverless applications and store the results in another service.
Often, stateless services implement common but computationally expensive tasks, such as image identification or speech parsing through services such as Amazon Rekognition~\cite{rekognition}.

\subsection{Serverless Function Composition}
\label{subsec:architectures}

There are three main ways that functions are composed into larger applications in serverless platforms. 

\paragraph{Direct Invocation.} Direct invocation is the simplest method of invoking successor serverless functions. A serverless function may directly invoke another serverless function by passing the identifier of the successor function into a library call that interacts with the serverless platform.

\paragraph{Composition Frameworks.} Most serverless platforms also implement frameworks for directly composing services and functions. These frameworks, such as Amazon StepFunctions~\cite{stepfunctions} and OpenWhisk Composer~\cite{whiskcomposer}, provide a declarative syntax for composing serverless functions and supported services. In addition to declaring the relationship between functions and services, these function composition frameworks also include higher level abstractions for program flows such as conditional branching based on the value of data returned by an earlier stage of the composition.

\begin{figure*}[t]
\hfill\includegraphics[width=\columnwidth]{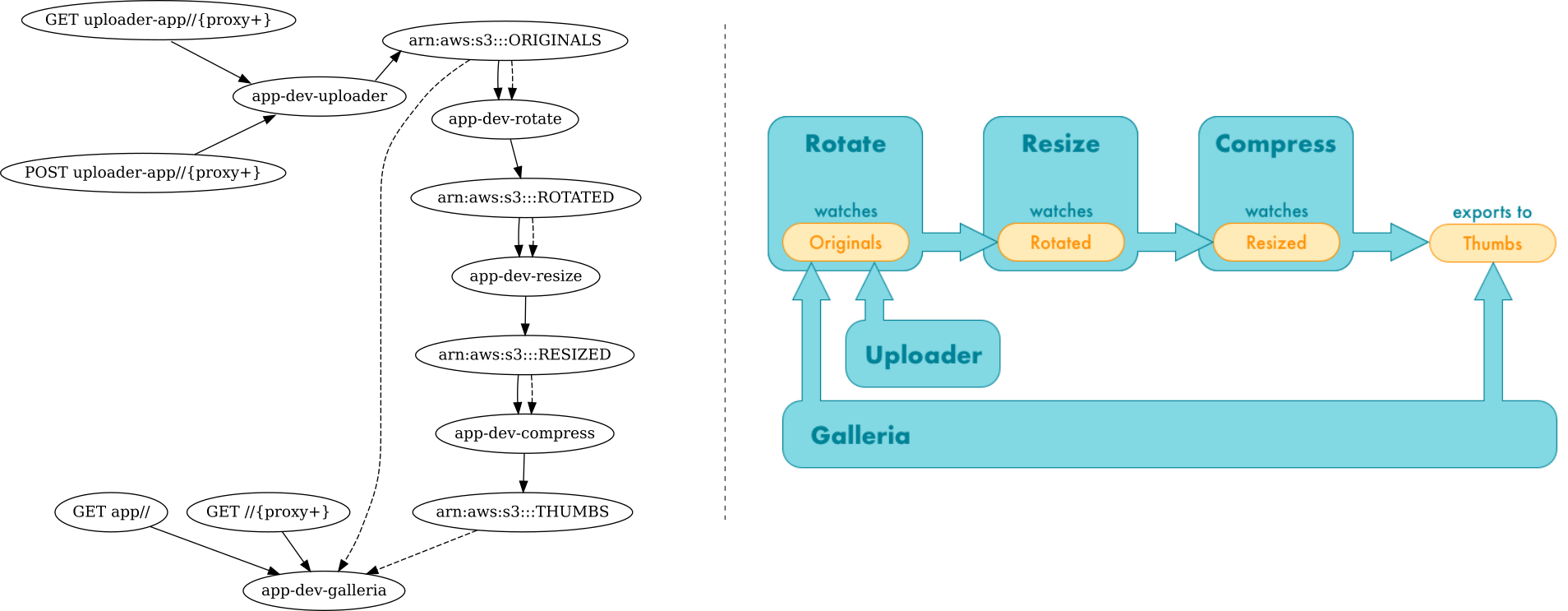}\hspace*{\fill}
\caption{Comparison of service call graph generated by our analysis for the galleria serverless application~\cite{galleria}, and pipeline diagram provided in the repository's user documentation. In the call graph at left, we see that GET and POST API gateway events in the top left of the call graph trigger the \texttt{app-dev-uploader} serverless function. This function then writes to the \texttt{ORIGINALS} S3 bucket, which in turn triggers the \texttt{app-dev-rotate} serverless function. This function reads from its triggering bucket then writes to a \texttt{ROTATED} bucket. The process repeats for two more image processing functions before the final image is uploaded to \texttt{THUMBS}.}
\label{fig:galleria-comparison}
\bottomfigureshrinker
\end{figure*}

\paragraph{Event Programming.} 
Serverless platforms provide a robust interface for specifying events that \textit{trigger} the execution of a function. Most platform services allow developers  to configure event triggers that activate when a service undergoes a state transition, e.g., when an object is created in an object store bucket. Upon being triggered, a serverless function will be provided with a copy of the data that triggered it, or an identifier to retrieve it from the associated service. 
In addition to events triggered by services, platforms also provide special gateways that handle interaction outside the platform, such as through HTTP requests.
We present a real-world example of function composition on the righthand-side of Figure~\ref{fig:galleria-comparison}. The sequence of functions performs image processing on uploaded files to generate consistently formatted thumbnails.


\vspace{-.2cm}
\section{Semantics for Serverless Computation}
\label{sec:semantics}

We introduce operational semantics for the execution of serverless applications. The goals of these serverless semantics are to: 1) precisely model the semantics of communication between serverless functions and platform services, and 2) capture program flows that are introduced as a result of this communication. 

\begin{figure*}[t]
	\resizebox{\textwidth}{!}{

\begin{tabular}{l|r}
		
		\begin{tabular}{lr}\
			$f \in F $ & defined functions \\
			$\sigma \in \Sigma $ & internal state \\
			$init \in F \times V \rightarrow \Sigma $ & initial state \\
			$v \coloneqq ...$ & value \\
			\hline \\
			$x \coloneqq ...$ & request ID \\
			$y \coloneqq ...$ & instance ID \\	
			$\mathbb{C} \coloneqq \mathbb{F}(f, \sigma, y)$ & executing serverless function \\
			$\phantom{C=}| \; \mathbb{R}(f,x,v)$ & received request \\
			$\phantom{C=}| \; \mathbb{S}(x,v)$ & generated response \\
			$step_f \in F \times \Sigma \rightarrow \Sigma$ & computational step \\
		\end{tabular}
		
		&
		
		\begin{tabular}{l}

			\LeftLabel{RECEIVE}
			\AxiomC{x is fresh}
			\UnaryInfC{$\mathbb{C} \Rightarrow \mathbb{C}\mathbb{R}(f,x,v)$}
			\DisplayProof 
			
			\vspace{2em} \\
			
			\LeftLabel{START}
			\AxiomC{}
			\UnaryInfC{$\mathbb{C}\mathbb{R}(f,x,v) \Rightarrow \mathbb{C}\mathbb{R}(f,x,v)\mathbb{F}(f,\mathit{init}(f,v),y)$}
			\DisplayProof
			
			\vspace{2em} \\
			
			\LeftLabel{COMPUTE}
			\AxiomC{$\mathit{step}_f(\sigma) = \sigma'$ }
			\UnaryInfC{$\mathbb{C}\mathbb{F}(f,\sigma, y) \Rightarrow \mathbb{C}\mathbb{F}(f, \sigma', y)$}
			\DisplayProof 
			
			\vspace{2em} \\
			
			\LeftLabel{RESPOND}
			\AxiomC{$\mathit{step}_f = \mathit{respond}(v')$}
			\UnaryInfC{$\mathbb{C}\mathbb{R}(f,x,v)\mathbb{F}(f, \sigma, y) \Rightarrow \mathbb{C}\mathbb{S}(x,v')\mathbb{F}(f, \sigma, y)$}
			\DisplayProof
			
			\vspace{2em} \\
			
			\LeftLabel{DIE}
			\AxiomC{}
			\UnaryInfC{$\mathbb{C}\mathbb{F}(f,\sigma, y) \Rightarrow \mathbb{C}$}
			\DisplayProof
			
		\end{tabular}
		
	\end{tabular}}
	\caption{In-process semantics models the sequence of steps in an individual serverless functions. A full serverless application $\mathbb{C}$ is modeled as a set of requests $\mathbb{R}$, executing functions $\mathbb{F}$, and generated responses $\mathbb{S}$. Functions and requests are appended to $\mathbb{C}$ as they become active, and are removed from $\mathbb{C}$ as they terminate or are responded to.  }
	\label{fig:process-semantics}
	\bottomfigureshrinker
\end{figure*}
\subsection{In-Process Semantics}

\textit{In-process semantics} for single serverless functions are defined in Figure~\ref{fig:process-semantics}. These semantics capture the sequence of steps in an individual serverless function. When an \emph{external} gateway service initiates a request for the execution of the serverless program, the platform applies the RECEIVE rule which adds a new request $\mathbb{R}$. The request contains a serverless function $f$ and a data value \textit{v} that will be passed to the function. Most commonly, RECEIVE represents a request made to a public web endpoint integrated with the serverless application. When an unhandled request exists, the platform applies the START rule which initializes $f$ with an initial state \textit{init(f,v)} and starts the execution of $f$. We note that \textit{init(f,v)} captures both initial state at cold and warm start. COMPUTE models the execution steps in a serverless function \textit{f}. Similarly to \cite{Jangda2019Formal}, COMPUTE is a language agnostic representation of transitions on state $\sigma$. COMPUTE absorbs interactions with platform services; Section~\ref{sec:event} details the rules for these interactions.  A serverless function may issue a response, in which case the platform applies the RESPOND rule. This rule will remove the unhandled request $\mathbb{R}(f,x,v)$ from the system and replace it with a response $\mathbb{S}(x,v')$, where $v'$ is a value provided by the RESPONDing serverless function. Responses represent data which is sent back to the external service that initiated the request; they are terminal states and are not used for further computation within the platform. Finally, functions may terminate through the application of the DIE rule. The system reaches a stable state when all requests have been responded to and no serverless functions are still executing.

\vspace{-.5cm}
\subsection{Event Semantics}
\label{sec:event}
We extend the in-process semantics with an event semantics to capture interaction of functions with platform services and direct invocation. 
We develop semantics for each service: object stores, databases, notifications, queues, and stateless services. These semantics detail how serverless functions interface with that specific service during execution. In this section we detail the semantics of object stores. We include the semantics for the remaining services in Appendix~\ref{sec:appendix}; these semantics follow the general structure of the object store semantics, however each details behavior specific to the service they model. 

The semantic rules can be broadly grouped into rules that \emph{write} the state of a service (UPLOAD and REMOVE for object stores; INSERT, UPDATE, and DELETE, for databases; and ENQUEUE for queues), and rules that \emph{read} data from a service into the state of an executing serverless function (READ for object stores, SELECT for databases, and DEQUEUE for queues).

\begin{figure}
\myfigureshrinker
	\begin{lstlisting}[language=yaml,basicstyle=\small]
functions:
  processor:
    handler: index.process
    events:
    - s3: 
      bucket: photos
      event: s3:ObjectCreated:*
	\end{lstlisting}
	      \vspace{-.3cm}	
	      \caption{An example event configuration. The serverless function \texttt{processor} is triggered when an object is added to the \texttt{photos} bucket. In the semantics, this event is represented as the fact $e(c, processor)$ where $c = (\mathit{photos},\mathit{upload})$. }
	\label{fig:eventdef}
	\bottomfigureshrinker
\end{figure}

Our semantics introduce a domain of events $E$ that captures function invocations due to service state transitions. An event $e(c,f) \in E$ consists of two parts: a triggering condition, $c$, and an associated serverless function $f$. Triggering conditions are generally defined by a unique service identifier $\mathit{sid}$ 
and an operation $\mathit{op}$ (e.g, $\mathit{upload}$ to an object store); we write $c = (\mathit{sid},\mathit{op})$. Program configurations unambiguously reference their associated services and the associated serverless functions. We reduce configurations to set of events $e(c,f)$ during static analysis. We present an example configuration in Figure~\ref{fig:eventdef}. An event is \emph{triggered} when a serverless function performs a step that fires the event condition. For instance, an upload to an object store $b$ will activate all events tied to upload to $b$. To capture the effect of these triggering events, our semantics introduce the function $\mathit{trigger}$. This function accepts a triggerring condition $c = (\mathit{sid},\mathit{op})$, and returns the set of functions $f$ for which there is $e(c,f)$, i.e., the set of functions that will execute when a function runs operation $\mathit{op}$ on service $\mathit{sid}$. We note that some types of triggering conditions defined in our semantics are officially supported by serverless platforms but rarely occur in practice, such as the trigger associated with a REMOVE from an object store.

Our semantics distinguishes between functions triggered by external requests and functions triggered by events on services. The platform applies RECEIVE followed by START on functions triggered by external requests. It immediately applies START on functions triggered by ``internal'' events on services. Our semantics allows that any function that is part of the serverless application may issue a response to the external request. RECEIVE and RESPOND define the ``boundary'' of the serverless 
application, although functions may continue to execute and modify services after a RESPOND.

\begin{figure*}
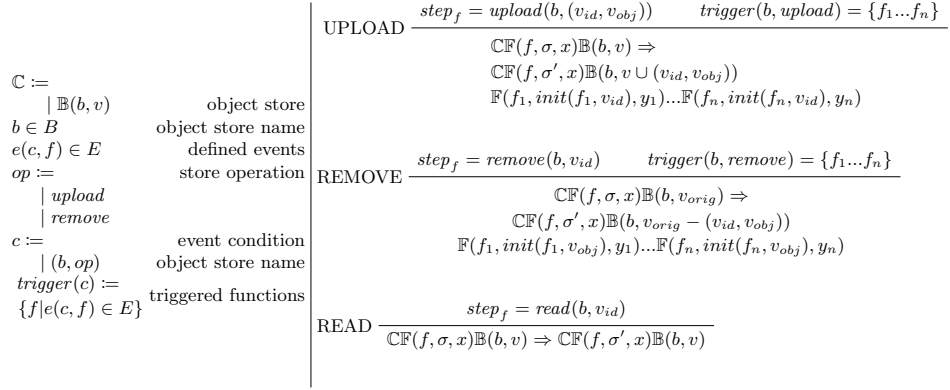

	\resizebox{\textwidth+2em}{!}{
\begin{tabular}{l|r}
	\begin{tabular}{lr}
		$\mathbb{C} \coloneqq $ & \\
		$\phantom{C=}| \; \mathbb{B}(b, v)$ & object store \\
		$b \in B$ & object store name \\
		$e(c,f) \in E$ & defined events \\
		$op \coloneqq$  & store operation \\
		$\phantom{c =} | \; \mathit{upload}$ & \\
		$\phantom{c =} | \; \mathit{remove}$ & \\
		$c \coloneqq$ & event condition \\
		$\phantom{c =} | \; (b,op)$ & object store name\\
		
		\stackanchor{$\mathit{trigger}(c) \coloneqq \quad$}{ $ \{ f | e(c,f) \in E \} $} & triggered functions
		
	\end{tabular}
&
	\begin{tabular}{l}
		
		\def\stackalignment{l}
		
		\LeftLabel{UPLOAD}
		\AxiomC{$\mathit{step}_f = \mathit{upload}(b, (v_{id}, v_{obj}))$}
		\AxiomC{$\mathit{trigger}(b, \mathit{upload}) = \{f_1...f_n\}$}
		\BinaryInfC{\Shortstack{
				{$\mathbb{C}\mathbb{F}(f,\sigma,x)\mathbb{B}(b,v) \Rightarrow$} {$\mathbb{C}\mathbb{F}(f,\sigma',x)\mathbb{B}(b,v \cup (v_{id}, v_{obj}))$}
				{$\mathbb{F}(f_1, init(f_1,v_{id}), y_1)...\mathbb{F}(f_n, init(f_n,v_{id}), y_n)$}
			}
		}
		\DisplayProof             
		
		\vspace{2em} \\

		\LeftLabel{REMOVE}
		\AxiomC{$\mathit{step}_f = \mathit{remove}(b, v_{id})$}
		\AxiomC{$\mathit{trigger}(b,\mathit{remove}) = \{f_1...f_n\}$}
		\BinaryInfC{\Shortstack{
				{$\mathbb{C}\mathbb{F}(f,\sigma,x)\mathbb{B}(b,v_{orig}) \Rightarrow$} {$\mathbb{C}\mathbb{F}(f,\sigma',x)\mathbb{B}(b,v_{orig} - (v_{id}, v_{obj}))$}
				{$\mathbb{F}(f_1, init(f_1,v_{obj}), y_1)...\mathbb{F}(f_n, init(f_n,v_{obj}), y_n)$}
			}
		}
		\DisplayProof                 
		
		\vspace{2em} \\
		
		\LeftLabel{READ}
		\AxiomC{$\mathit{step}_f = \mathit{read}(b, v_{id})$}
		\UnaryInfC{$\mathbb{C}\mathbb{F}(f,\sigma,x)\mathbb{B}(b,v) \Rightarrow \mathbb{C}\mathbb{F}(f,\sigma',x)\mathbb{B}(b,v)$}
		\DisplayProof                 
		
		\vspace{2em} \\
		
	\end{tabular}
\end{tabular}
}
	\caption{Object store event semantics.}
	\label{fig:store-semantics}
	\bottomfigureshrinker
\end{figure*}

We define semantics for object stores in Figure~\ref{fig:store-semantics}. Each object store has a unique identifier $b$ in $B$, the set of object stores defined for the application. Object stores provide a filesystem-like interface for writing and reading data. In the semantics, this interaction is encoded by allowing serverless functions to write or overwrite some value $v$ in a named bucket by applying the UPLOAD rule. When a file is uploaded, all events triggered by state transition on the receiving bucket initialize their respective function(s). Serverless functions can also delete data contained in a bucket through application of the REMOVE rule. When a function retrieves a data value from a bucket, 
the READ rule accesses the associated data and assigns it to a variable inside the function's local state.

 Our event semantics are synchronous in the sense that a request to a service and the execution of the request by the service happen in ``one step''. This facilitates static reasoning. In practice, a request is decoupled from the execution; we conjecture that the synchronous semantics are sufficient as programs implicitly synchronize events on services: a read in $f_2$ \emph{is triggered} by a write in $f_1$. 
Further, for reads and writes within the same function, standard libraries typically provide only synchronous methods for interacting with platform services.
We will formalize sufficiency conditions on programs in future work.

\vspace{-.2cm}
\subsection{Platform Behavior Encoded in Semantics}
\vspace{-.1cm}
Our semantics are sufficiently expressive to capture features of serverless platforms that impact system state in unintuitive ways. The non-finality of RESPOND and the effects caused by function retries are two examples of this behavior, which we discuss below.

\begin{figure}[t]
	\lstset{language=JavaScript,
		breaklines=false,
		 basicstyle=\small
	}
	\begin{lstlisting}
export.shortenUrl = function(event, context, callback) {
  let url = event.body;
  let slug = crypto.randomBytes(8).toString(...).replace(...);
  callback(null, {shortUrl: context.domainName + slug});
  dynamodb.put({
    TableName: "ShortUrls",
    Item: {slug: slug,long_url: url}
  });
}
	\end{lstlisting}
	\vspace{-.4cm}
	\caption{Example of execution continuing after response. RESPOND is applied when the \texttt{callback} passed in to the serverless function is invoked, but a database is written to after this response. Code adapted from the \texttt{url-shortener} project~\cite{urlshortener}.}
	\label{fig:earlycallback}	
\bottomfigureshrinker
\end{figure}

\vspace{-.2cm}
\subsubsection{Non-finality of RESPOND.}  Unlike \texttt{return} statements in normal functions, responses from a serverless function do not return from the function. Consider the example in Figure~\ref{fig:earlycallback}. This serverless function accepts a URL string and generates a random short slug for that URL. It immediately responds with the generated shortened URL, then afterward writes the association between the slug and the original URL to a database. Our semantics models the execution of this serverless function by the following transitions in our semantics ($\mathbb{D}$ represents the database service. INSERT has semantics similar to UPLOAD in Figure~\ref{fig:store-semantics}):
\small \begin{align*} 
& \mathbb{C}\mathbb{D}(\mathit{ShortUrls}, v) & \quad & \text{} \\
\implies & \mathbb{C}\mathbb{D}(\mathit{ShortUrls}, v)\mathbb{R}(f, x, v_1) & \quad & \text{by rule RECEIVE($f$, $x$, $v_1$)} \\
\implies & \mathbb{C}\mathbb{D}(\mathit{ShortUrls}, v)\mathbb{R}(f, x, v_1)\mathbb{F}(f, \sigma, y) & \quad &  \text{by START($y$)}\\
\implies & \mathbb{C}\mathbb{D}(\mathit{ShortUrls}, v)\mathbb{R}(f, x, v_1) & & \\
         & \phantom{\mathbb{C}}\mathbb{F}(f, \sigma' = \sigma[url\leftarrow ev.body,slug\leftarrow rand()], y) & \quad &  \text{by COMPUTE($f$)} \\
\implies & \mathbb{C}\mathbb{D}(\mathit{ShortUrls}, v)\mathbb{S}(x, v')\mathbb{F}(f, \sigma', y) & \quad &  \text{by RESPOND($x$, $v'=\sigma'[slug]$)} \\
\implies & \mathbb{C}\mathbb{D}(\mathit{ShortUrls}, v' \cup v)\mathbb{S}(x, v')\mathbb{F}(f, \sigma', y) & \quad &  \text{by INSERT($\mathit{ShortUrls}$, $v'$)} \\
\implies & \mathbb{C}\mathbb{D}(\mathit{ShortUrls}, v' \cup v)\mathbb{S}(x, v')& \quad &  \text{by DIE($y$)} \\
\end{align*} \normalsize
The application of the INSERT rule affects the final state of the system $\mathbb{C}$ by introducing the value $v'$ to the database $\mathbb{D}$. This insertion occurs even though the serverless function has already generated a response in an earlier step. 

\subsubsection{Failures and Retried Executions.} 
A serverless function may fail during execution for two reasons 1) the function code has entered an error state as the result of an uncaught exception, or 2) the container runtime has killed the function, either because execution has timed out, or because the language interpreter has failed with an error. When a  function fails, the platform can retry the function by starting a new execution with a clone of the data from the original request~\cite{awslambda}. 

Our semantics capture the effects of failures and retried executions that may impact system state. 
In particular, serverless functions that are not idempotent may emit messages to platform services that are repeated in retried executions, affecting final system state. In our semantics, these retries are modeled as an application of the DIE rule, followed by a subsequent application of START to handle a still-unsatisfied request. Consider a serverless function that uses the UPDATE rule to increment a view count. It is retried due to a spontaneous failure in the data center where the function is executing. This series of events are modeled under our semantics as:
\small \begin{align*}
& \mathbb{C}\mathbb{D}(\mathit{ViewCount}, v) & & \text{} \\
\implies & \mathbb{C}\mathbb{D}(\mathit{ViewCount}, v)\mathbb{R}(f, x, v) & & \text{by RECEIVE($f$, $x$, $v$)} \\
\implies & \mathbb{C}\mathbb{D}(\mathit{ViewCount}, v)\mathbb{R}(f, x, v) \mathbb{F}(f, \sigma, y) & &  \text{by START($y$)}\\
\implies & \mathbb{C}\mathbb{D}(\mathit{ViewCount}, v+1)\mathbb{R}(f, x, v) \mathbb{F}(f, \sigma, y) &  &  \text{by UPDATE($\mathit{ViewCount}$, $(v)$$\rightarrow$ $v+1$)} \\
\implies & \mathbb{C}\mathbb{D}(\mathit{ViewCount}, v+1)\mathbb{R}(f, x, v)  & &  \text{by DIE($y$)} \\
\implies & \mathbb{C}\mathbb{D}(\mathit{ViewCount}, v+1)\mathbb{R}(f, x, v) \mathbb{F}(f, \sigma, y') &  &  \text{by START($y'$)} \\
\implies & \mathbb{C}\mathbb{D}(\mathit{ViewCount}, v+2)\mathbb{R}(f, x, v) \mathbb{F}(f, \sigma, y') &  &  \text{by UPDATE($\mathit{ViewCount}$, $(v)$$\rightarrow$ $v+1$)} \\
\implies & \mathbb{C}\mathbb{D}(\mathit{ViewCount}, v+2)\mathbb{S}(x, v) \mathbb{F}(f, \sigma, y') &  &  \text{by RESPOND($x$, \{\})} \\
\implies & \mathbb{C}\mathbb{D}(\mathit{ViewCount}, v+2)\mathbb{S}(x, \{\}) &  &  \text{by DIE($y'$)} 
\end{align*}
\normalsize
We observe that following these state transitions, the $\mathit{ViewCount}$ of the database has been incremented twice, despite only a single request being made to the serverless function. Such faults are representative of data inconsistencies that exist in real serverless applications that violate the idempotency recommended by serverless providers~\cite{awslambda}.

\subsection{Platform Supported Function Composition}  
\label{sec:stepfunctions}
Function composition frameworks allow developers to statically declare pathways for messages through a serverless application. When one of these pathways is defined, the return value of a serverless function implicitly becomes a message passed to the serverless function or service following it in the composition. This occurs without the explicit invocation of a library method to cause state transfer on a platform service that is used for other serverless events. Despite this difference, our semantics are expressive enough to capture such behavior using the same set of state transitions as other serverless events.

Consider a StepFunction composition that defines a chain of two serverless functions, $f_1$ and $f_2$. (The Appendix~\ref{sec:appendix} provides an example StepFunction declaration.) Our semantics models the execution as follows:
\vspace{-1em}
\small
\begin{align*}
& \mathbb{C}& \qquad & \text{} \\
\implies & \mathbb{C}\mathbb{R}(f_{\mathit{step}}, x, v) & \qquad & \text{by RECEIVE($f_{\mathit{step}}$, $x$, $v$)} \\
\implies & \mathbb{C}\mathbb{R}(f_{\mathit{step}}, x, v) \mathbb{F}(f_1, \sigma, y) & \qquad &  \text{by START($y$)}\\
\implies & \mathbb{C}\mathbb{R}(f_{\mathit{step}}, x, v) \mathbb{F}(f_1, \sigma', y) & \qquad &  \text{by COMPUTE($f_1$)} \\
\implies & \mathbb{C}\mathbb{R}(f_{\mathit{step}}, x, v) \mathbb{F}(f_1, \sigma', y)\mathbb{F}(f_2, \sigma, y)  & \qquad &  \text{by INVOKE($f_2$, $v'$)} \\
\implies & \mathbb{C}\mathbb{R}(f_{\mathit{step}}, x, v) \mathbb{F}(f_2, \sigma, y)  & \qquad &  \text{by DIE($f_1$, $\sigma$, $y$)} \\
\implies & \mathbb{C}\mathbb{R}(f_{\mathit{step}}, x, v) \mathbb{F}(f_2, \sigma', y) & \qquad &  \text{by COMPUTE($f_2$)} \\
\implies & \mathbb{C}\mathbb{S}(x, v'') \mathbb{F}(f_2, \sigma'', y) & \qquad &  \text{by RESPOND($x$, $v''$)} \\
\implies & \mathbb{C}\mathbb{S}(x, v'')& \qquad &  \text{by DIE($f_2$, $\sigma''$, $y$)} 
\end{align*}
\normalsize

This execution illustrates an important difference between standalone serverless functions and those defined as part of a composition chain. The platform
starts the StepFunction chain by issuing a request $\mathbb{R}(f_{\mathit{step}}, x, v)$ by RECEIVE.
Only the final serverless function in the chain RESPONDs to the request. The ``return'' of all other functions in the chain is encoded as an event rule that activates the next function in the chain. Since compositions are static, the target of each stage of the composition is known. To preserve the connection to the originating StepFunction request that started $\mathbb{F}(f_1,\sigma,y)$, $f_2$ inherits the identifier $y$ from $f_1$ when it is invoked. Thus, the lifecycle of the first function in the composition chain is RECEIVE, START, COMPUTE, DIE; the lifecycle of the final one is COMPUTE, RESPOND, DIE.


\vspace{-.2cm}
\section{Service Call Graphs}
\label{sec:callgraph}
\begin{figure}[t]
\myfigureshrinker
\centering
\includegraphics[width=0.4\columnwidth]{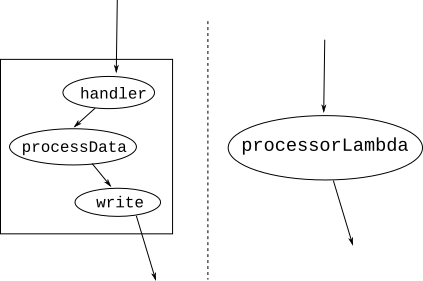}
	\caption{Example simplification of a serverless function node. When \texttt{processorLambda} is invoked, the platform executes its \textit{handler}. The handler may be split into several local helper functions, which we represent as a single node, shown at right.}
	\label{fig:lambdasimp}
	\bottomfigureshrinker
\end{figure}

\vspace{-.1cm}
Our semantics enable construction of a \emph{service call graph} that explicitly models interaction between services and serverless functions. The service call graph extends the classical call graph by adding nodes that represent platform services and edges that represent reads from services, writes to services, and transfer of control to functions triggered by state transition on services. We simplify our graphs by treating an entire intra-function call graph as a single node as demonstrated in Figure~\ref{fig:lambdasimp} to more clearly capture the interaction \textit{between} functions and platform services.

Construction of the service call graph proceeds in two phases: \emph{configuration analysis}, and \emph{code analysis}. Configuration analysis processes configuration files and identifies the serverless functions and services for the given application. Each serverless function $f \in F$, and each service $b \in B$ (object store), $d \in D$ (database), $q \in Q$ (queue), and $t \in T$ (notification topic) becomes a node in the service call graph. In addition to identifying the set of functions and services, configuration analysis also identifies the set of events $e(c,f) \in E$ and triggering conditions $c=(\mathit{sid},\mathit{op})$ (recall Section~\ref{sec:semantics} for the explanation on $e$ and $c$); each event $e(c,f) \in E$ where $c=(\mathit{sid},\mathit{op})$ gives rise to an edge from $\mathit{sid}$ to $f$. 

Code analysis processes each serverless functions $f \in F$. It constructs the standard interprocedural control flow graph (ICFG) of $f$ (we note that here ``interprocedural'' refers to the intra-function call graph of $f$). The analysis tracks the set of service identifiers $\mathit{sid}$ that flow to call sites in the ICFG corresponding to rules of the event semantics (such as UPLOAD, ENQUEUE, INSERT, or NOTIFY). At each such call site, the analysis adds an edge from the current serverless function $f$ to each service $\mathit{sid}$ that may reach the call site corresponding to the event rule. For instance, consider the execution of serverless function $f_1$ captured in the semantics as:
\small
\begin{align*}
 & \mathbb{C}\mathbb{D}(\mathit{someTable}, v)\mathbb{F}(f_1, \sigma, x) & &  \text{}\\
\implies & \mathbb{C}\mathbb{D}(\mathit{someTable}, v)\mathbb{F}(f_1, \sigma' = \sigma[id \leftarrow \mathit{someTable}], x) & &  \text{by COMPUTE}\\
\implies & \mathbb{C}\mathbb{D}(\mathit{someTable}, v)\mathbb{F}(f_1, \sigma'' = \sigma'[data \leftarrow \mathit{v_{new}}], x) & &  \text{by COMPUTE}\\
\implies & \mathbb{C}\mathbb{D}(\mathit{someTable}, v \cup \mathit{v_{new}}) & &\\
         & \phantom{\mathbb{C}}\mathbb{F}(f_1, \sigma'', x)\mathbb{F}(f_2, \mathit{init}(f_2, \mathit{v_{new}}), y) & &  \text{by INSERT($\sigma''[\mathit{id}]$, $\sigma''[\mathit{data}]$)} \\
\end{align*}
\normalsize
In this function, identifier \textit{someTable} is assigned to the variable \textit{id} as a step in the execution of $f_1$. When the $f_1$ later applies INSERT, all possible values of \textit{id} flow to the parameter of INSERT. Since \textit{id} can only have the value of \textit{someTable} at this time, the analysis adds an edge in the service call graph from $f_1$ to the service $someTable$. Since there is a triggering event $e(c,f_2) \in E$ where $c = (\textit{someTable}$, INSERT$)$ in the configuration, 
this INSERT also triggers the execution of $f_2$.

There are a variety of applications of our semantics and static service call graphs~\cite{berkeley2019cloud}. One application is \emph{container prewarming}; the platform can use the call graph to prepare containers for functions scheduled for execution and reduce the penalty of cold starts. In our experiments, warm starts reduced running time for \texttt{galleria} in Figure~\ref{fig:galleria-comparison} by approx. 25\% (Appendix~\ref{sec:appendix}).
Other applications include resource usage prediction, information flow analysis, and static debugging.


\vspace{-.3cm}
\section{Call Graph Implementation and Evaluation}\label{sec:implementation}
\vspace{-.2cm}

We implement service call graph construction as an extension of the Type Analysis for JavaScript framework~\cite{tajs2009}. Specifically, we employ a branch of TAJS that supports reasoning about asynchronous behavior~\cite{sotiropoulos2019static}.
Our analysis consists of 1187 lines of Java code that interface with the TAJS intermediate representation of JavaScript to build the service call graph, and 245 lines of JavaScript code that summarize the effects of third party libraries, including the AWS SDK. 
We constructed summaries of library functions to overcome limitations in TAJS that prevented us from performing standard whole-program analysis. 
We intend to release our implementation publicly. 

We generate call graphs for applications collected from GitHub. We searched GitHub for repositories that included serverless configuration files that defined more than one serverless function, sorted by repository popularity. We analyze the top twelve applications that fit this criteria. 
To evaluate the accuracy of our generated call graphs, we compare the output of our analysis against call graphs drawn by manual inspection of programs.

\begin{table*}[t]\centering
	\caption{Service Call Graph results.}
	\scriptsize
	\begin{tabular}{|l|r|r|c|r|} 
		\hline
		\multicolumn{1}{|p{10em}|}{ \centering Application}
		& \multicolumn{1}{|p{5em}|}{\centering  Lines of Code }
		& \multicolumn{1}{|p{5em}|}{\centering  \# Functions}
		& \multicolumn{1}{|p{5em}|}{\centering  Sound? }
		& \multicolumn{1}{|p{7em}|}{\centering  Missed Edges} \\
		\hline
		hello-retail &  2288 & 14 & Y & 0 \\
		\hline
		citizen-dispatch & 865 & 3 & N & 6 \\
		\hline
		galleria &  641 & 5 & Y & 0 \\
		\hline
		rating-service & 412 & 2 & Y & 0 \\
		\hline
		LEX & 323 & 2 & Y & 0 \\
		\hline
		lending-app & 258 & 4 & Y & 0\\
		\hline
		url-shortener & 172 & 3 & Y & 0 \\
		\hline
		zen-beer & 155 & 4 & Y & 0 \\
		\hline
		greeting-app & 99 & 2  & Y & 0\\
		\hline
		lane-breach & 98 & 2 & N & 1 \\
		\hline
		wombat & 88 & 2 & Y & 0 \\
		\hline
		serverless-chaining & 28 & 2 & Y & 0 \\
		\hline
	\end{tabular}
	\label{tab:codemetrics} 
\end{table*}

Table~\ref{tab:codemetrics} presents the analysis results. For 10 of the 12 applications, our analysis produced a service call graph \emph{identical to the ground truth}.
One such comparison is shown in Figure~\ref{fig:galleria-comparison}.
For two applications, our analysis missed edges. In the case of \texttt{lane-breach}, the missed edge corresponded to a web request made directly to another function through the external web API. We note that it is not possible, in general, to determine whether a web address belongs to the application under analysis or a third-party web site. 
Fortunately, this behavior represents a discouraged pattern~\cite{baldini2017serverless}; the program could be made more efficient by rewriting the code to use a direct invocation, which would be captured through our INVOKE rule. 

In the case of \texttt{citizen-dispatch}, the analysis missed edges from serverless functions to a set of database tables that corresponded to database queries made by third-party library calls. This program violated our assumption that third-party libraries do not interact with services.
Though constant service identifiers flow to the library calls, it is difficult to statically infer which tables will be accessed by a particular call due to the nature of the query inference engine. Future versions of our tool could safely over-approximate this behavior by assuming that any library call
 has the potential to query all tables. If we could perform standard whole-program analysis, interactions with the database through the library would have been soundly detected. (Whole-program analysis is trivially supported in tools for languages such as Java, but is not supported by TAJS due to the difficulty of analyzing JavaScript.)


\vspace{-.3cm}
\section{Conclusion}
\vspace{-.2cm}
\label{sec:conclusion}

We have introduced new operational semantics for serverless computing. 
We have demonstrated how these semantics can be used to produce a new type of call graph that incorporates services and event-dependent program flows.
Finally, we have presented a prototype of our call graph construction algorithm and showed its efficacy on real-world serverless programs.
In future work, we will use these semantics to construct analyses and tools for improving performance and security of serverless applications.

%
%
\bibliographystyle{splncs04}

\newpage

\appendix



\section{Semantics of Serverless Computation} \label{sec:appendix}

\begin{figure*}
	\resizebox{\textwidth+2em}{!}{
	\begin{tabular}{l|r}
	\begin{tabular}{lr}
		$\mathbb{C} \coloneqq $ & \\
		$\phantom{C=}| \; \mathbb{D}(d, v)$ & database service \\
		$d \in D$ & database table name \\
		$op  \coloneqq $ & database operation \\
		$\phantom{op=} | \; insert$ & database insertion \\
		$\phantom{op=} | \; delete$ & database deletion \\
		$\phantom{op=} | \; update$ & database update \\
		$query \in V \rightarrow V$ & effect of query \\
		$e(c,f) \in E$ & defined events \\
		$c \coloneqq$ & event condition \\
		$\phantom{c =} | \;(d, op)$ & database query \\
		
		\stackanchor{$trigger(c) \in c \rightarrow F$}{ $\coloneqq \{ f | e(c,f) \in E \} $} & triggered functions
		
	\end{tabular}
&
	\begin{tabular}{l}
		
		\def\stackalignment{l}		
		\LeftLabel{INSERT}
		\AxiomC{$step_f=insert(d,v_{new})$}
		\AxiomC{$\mathit{trigger}((d, insert))= \{f_1...f_n\}$}
		\BinaryInfC{\Shortstack{
				{$\mathbb{C}\mathbb{F}(f,\sigma, x)\mathbb{D}(d,v) \Rightarrow$}
				{$\mathbb{C}\mathbb{F}(f,\sigma', x)\mathbb{D}(d,v \cup v_{new})$}
				{$\mathbb{F}(f_1, init(f_1,v_{new}), y_1)...\mathbb{F}(f_n, init(f_n,v_{new}), y_n)$}
			}
		}
		\DisplayProof 
		
		\vspace{2em} \\
		
		\LeftLabel{UPDATE}
		\AxiomC{$step_f=update(d,query)$}
		\AxiomC{$trigger((d, update))= \{f_1...f_n\}$}
		\BinaryInfC{\Shortstack{
				{$\mathbb{C}\mathbb{F}(f,\sigma, x)\mathbb{D}(d,v) \Rightarrow$}
				{ $\mathbb{C}\mathbb{F}(f,\sigma', x)\mathbb{D}(d,query(v))$}
				{$\mathbb{F}(f_1, init(f_1,v_{\mathit{modified}}), y_1)...\mathbb{F}(f_n, init(f_n,v_{\mathit{modified}}), y_n)$}
			}
		}
		\DisplayProof 
		
		\vspace{2em} \\
		
		\LeftLabel{DELETE}
		\AxiomC{$step_f=delete(d,query)$}
		\UnaryInfC{$\mathbb{C}\mathbb{F}(f,\sigma,x)\mathbb{D}(d,v) \Rightarrow \mathbb{C}\mathbb{F}(f,\sigma',x)\mathbb{D}(d,query(v))$}
		\DisplayProof                 
		
		\vspace{2em} \\
		
		\LeftLabel{SELECT}
		\AxiomC{$step_f=select(d,query)$}
		\UnaryInfC{$\mathbb{C}\mathbb{F}(f,\sigma,x)\mathbb{D}(d,v) \Rightarrow \mathbb{C}\mathbb{F}(f,\sigma',x)\mathbb{D}(d,v)$}
		\DisplayProof                 
		
	\end{tabular}
\end{tabular}
	}
	\caption{Database event semantics.}
	\label{fig:database-semantics}
\end{figure*}

\subsubsection{Databases.} The database semantics in Figure~\ref{fig:database-semantics} are similar to object stores. As with object stores, each table of a database has a uniquely identifying table name $d$ in $D$, the global domain of database tables defined for a serverless applications. Serverless functions may trigger other functions by adding data to a database using the INSERT or UPDATE rules. They may also remove existing data using the DELETE rule and access data with the SELECT rule. However, unlike object stores, databases allow for complex queries which may operate on several values in a single step. In order to encapsulate the effect of database queries, we define the function $\mathit{query}(v)$, which accepts a database value $v$ and produces some resulting value $v'$. When a serverless function performs a step that acts on a database, the step receives as input a query function that is used to compute the state transfer on a database and select returned rows. This abstraction allows the effects of database querying to be reasoned about without the need to define semantics for the relational algebra operations supported by serverless databases.

\subsubsection{Serverless Queues.} The queue semantics defined in Figure~\ref{fig:queue-semantics} are distinct from other platform services in that data cannot be read from a queue into a currently executing serverless function. Instead, each individual queue $q$ in $Q$, the global domain of queues defined for a serverless application, acts as a buffer for data that will be processed by new invocations of serverless functions. Serverless functions may append data to a queue by applying the ENQUEUE rule. When a serverless platform detects that a queue meets service-specific conditions, it pops data from that queue using the DEQUEUE rule and passes it as a parameter into a new instance of each serverless function that is triggered by that queue.

\begin{figure*}
	\resizebox{\textwidth+2em}{!}{
	\begin{tabular}{l|r}
	\begin{tabular}{lr}
		$\mathbb{C} \coloneqq $ & \\
		
		$\phantom{C=}| \; \mathbb{Q}(q, v)$ & queue service \\
		$q \in Q$ & queue name \\
		$e(c,f) \in E$ & defined events \\
		$c \coloneqq$ & event condition \\
		$\phantom{c =} | \; q$ & queue name\\
		$ready(r) \rightarrow bool$ & queue triggering condition \\
		\stackanchor{$trigger(c) \in c \rightarrow F$}{ $\coloneqq \{ f | e(c,f) \in E \} $} & triggered functions

	\end{tabular}
&
	\begin{tabular}{l}
		
		\def\stackalignment{l}
		
		\LeftLabel{ENQUEUE}
		\AxiomC{$step_f=enqueue(q,v_{new})$}
		\UnaryInfC{$\mathbb{C}\mathbb{F}(f,\sigma,x)\mathbb{Q}(q,v) \Rightarrow\mathbb{C}\mathbb{F}(f,\sigma',x)\mathbb{Q}(q,[v;v_{new}])$}
		\DisplayProof                 
		
		\vspace{2em} \\

		\LeftLabel{DEQUEUE}
		\AxiomC{$\mathit{ready(q)}$ \quad $trigger(q) = \{f_1...f_n\}$}
		\UnaryInfC{\Shortstack{
				{$\mathbb{C}\mathbb{Q}(q,[v_1;v_{remain}]) \Rightarrow \mathbb{C}\mathbb{Q}(q,v_{remain})$}
				{$\mathbb{F}(f_1, init(f_1,v_1), y_1)...\mathbb{F}(f_n, init(f_n,v_1), y_n)$}
			}
		}
		\DisplayProof      
		
	\end{tabular}
\end{tabular}
	}
	\caption{Queue event semantics.}
	\label{fig:queue-semantics}
\end{figure*}

\subsubsection{Stateless Services.} 
Our semantics also support stateless services through rules defined in Figure~\ref{fig:stateless-semantics}. We encode interactions with stateless services through the SERVICE rule of our event semantics. In this rule, an invocation of a stateless service is provided with data $v$ as well as an event condition $c$. The event condition $c$ serves as the identifier for the service where the stateless service should externally store the result of its computation on $v$. This write to $c$ by the stateless service will cause functions with events triggered by writes to $c$ through $\mathit{trigger(c)}$ to execute as normal. Additionally, we encode the behavior of stateless notification services through the NOTIFY rule. When a serverless function publishes data $v$ to some topic $t$ in $T$, the global set of defined topics, all functions which subscribe to the topic $t$ are triggered.  In addition to communication between functions and services, functions can also directly invoke other functions as a step of the function body. We represent this behavior through the INVOKE rule.

\begin{figure*}
	\resizebox{\textwidth+2em}{!}{
	\begin{tabular}{l|r}
	\begin{tabular}{lr}
		$t \in T$ & notification topic \\
		$e(c,f) \in E$ & defined events \\
		$c \coloneqq$ & event condition \\
		$\phantom{c=} | \; t$ & notification topic\\
		$res \in V \rightarrow V$ & effect of external service\\
		
		\stackanchor{$trigger(c) \in c \rightarrow F$}{ $\coloneqq \{ f | e(c,f) \in E \} $} & triggered functions
		
	\end{tabular}
&
	\begin{tabular}{l}
		
		\def\stackalignment{l}

		\LeftLabel{NOTIFY}
		\AxiomC{$step_f = \mathit{notify}(t, v)$}
		\AxiomC{ $trigger(t) = \{f_1...f_n\}$}
		\BinaryInfC{\Shortstack{
				{$\mathbb{C}\mathbb{F}(f,\sigma, x) \Rightarrow \mathbb{C}\mathbb{F}(f,\sigma', x)$}
				{$\mathbb{F}(f_1, init(f_1,v), y_1)...\mathbb{F}(f_n, init(f_n,v), y_n)$}
			}
		}
		\DisplayProof 
		
		\vspace{2em} \\
		
		\LeftLabel{SERVICE}
		\AxiomC{$step_f = service(v,c)$}
		\AxiomC{$trigger(c) = \{f_1...f_n\}$}
		\BinaryInfC{
			\Shortstack{{$\mathbb{C}\mathbb{F}(f,\sigma,x) \Rightarrow \mathbb{C}\mathbb{F}(f,\sigma',x)$}
				{$\mathbb{F}(f_1, init(f_1,res(v)), y_1)...\mathbb{F}(f_n, init(f_n,res(v)), y_n)$}
			}
		}
		\DisplayProof
		\vspace{2em} \\
		
		\LeftLabel{INVOKE}
		\AxiomC{$step_f = invoke(f_{new},v)$}
		\UnaryInfC{$\mathbb{C}\mathbb{F}(f,\sigma,x) \Rightarrow \mathbb{C}\mathbb{F}(f,\sigma',x)\mathbb{F}(f_{new},init(f_{new},v),y)$}
		\DisplayProof
	\end{tabular}
\end{tabular}
	}
	\caption{Stateless service event semantics. }
	\label{fig:stateless-semantics}
\end{figure*}

\begin{figure}
\begin{lstlisting}[language=yaml]
receive:
 Type: Task
 Resource: "arn:aws:lambda:us-east-1:XXX:function:receive"
 Next: parallel
parallel:
 Type: Parallel
 ResultPath: \$.results.parallel
 Branches:
 - StartAt: log
    States:
      log:
        Type: Task
        Resource: "arn:aws:lambda:us-east-1:XXX:function:log"
        End: true
 - StartAt: auth
    States:
      auth:
        Type: Task
        Resource: "arn:aws:lambda:us-east-1:XXX:function:auth"
        End: true
        ResultPath: \$.results.authorize	
\end{lstlisting}
	\caption{Example of parallelism in a StepFunction. In this composition, the \texttt{receive} serverless function will INVOKE both \texttt{log} and \texttt{auth}.}
	\label{fig:parallel}
\end{figure}

\subsubsection{Platform Supported Function Composition: Composition Parallelism and Conditionals.} Platform supported composition allows multiple functions to be executed in response to a single event. We provide an example of such a configuration in Figure~\ref{fig:parallel}. In our semantics, such parallelism is encoded by applying the necessary rules repeatedly, once for each starting point in the parallel section of the composition. For instance, the parallel portion of the execution of Figure~\ref{fig:parallel} would be encoded as:
 
 \small
 \begin{align*}
 & \mathbb{C}\mathbb{R}(f_{step}, x, v) \mathbb{F}(f_{\mathit{recv}}, \sigma, y) & \qquad & \text{} \\
 \implies & \mathbb{C}\mathbb{R}(f_{\mathit{step}}, x, v) \mathbb{F}(f_{\mathit{recv}}, \sigma, y)\mathbb{F}(f_{\mathit{log}}, \sigma, y)  & \qquad &  \text{INVOKE($f_{\mathit{log}}$, $v'$)} \\
 \implies & \mathbb{C}\mathbb{R}(f_{\mathit{step}}, x, v) \mathbb{F}(f_{\mathit{recv}}, \sigma, y)\mathbb{F}(f_{\mathit{log}}, \sigma, y)\mathbb{F}(f_{\mathit{auth}}, \sigma, y)  & \qquad &  \text{INVOKE($f_{\mathit{auth}}$, $v'$)} \\
 \implies & \mathbb{C}\mathbb{R}(f_{\mathit{step}}, x, v) \mathbb{F}(f_{\mathit{recv}}, \sigma, y)\mathbb{F}(f_{\mathit{auth}}, \sigma, y)  & \qquad &  \text{DIE($f_{\mathit{log}}$)} \\
 \implies & \mathbb{C}\mathbb{R}(f_{\mathit{step}}, x, v) \mathbb{F}(f_{\mathit{recv}}, \sigma, y)  & \qquad &  \text{DIE($f_{\mathit{auth}}$)} \\
 \implies & \mathbb{C}\mathbb{S}(x,v') \mathbb{F}(f_{\mathit{recv}}, \sigma, y)  & \qquad &  \text{RESPOND($x$, $v'$)} \\
 \end{align*}
 \normalsize
 
Our semantics assumes that the function that spawns the parallel arms acts as a barrier. In the above execution \textit{recv} joins \textit{log} and \textit{auth}, then it RESPONDs to the StepFunction request. 

\begin{figure}
\myfigureshrinker
	\lstinputlisting[language=yaml]{
	simple-state-machine.tex}
	\caption{Example of platform supported function composition using AWS StepFunctions. In this example, a web request triggers the \texttt{RecordDB} serverless function. When \texttt{RecordDB} completes, it INVOKES the serverless function \texttt{RecordAC}. Code is modified from the \texttt{slack-signup-serverless} project~\cite{slacksignup}.}
	\label{fig:simple-state-machine}
\end{figure}

\begin{figure}
\begin{lstlisting}[language=yaml]
AuthOrNot:
  Type: Choice
  Choices:
   - Variable: "\$.results.receive.doAuth"
       NumericEquals: 1
       Next: authorize
  Default: fail
authorize:
  Type: Task
  Resource: "arn:aws:lambda:us-east-1:XXX:function:auth"
  End: true
  ResultPath: \$.results.authorize	
fail:
  Type: Task
  Resource: "arn:aws:lambda:us-east-1:XXX:function:fail"
  End: true
  ResultPath: \$.results.respond
\end{lstlisting}
	\caption{Example of branching behavior in StepFunctions. In this example, the \texttt{AuthOrNot} step evaluates the value of a \texttt{doAuth} field against the numeric literals 1 and 0. In the case of 1 it executes the \texttt{authorize} function, otherwise it executes the \texttt{fail} function.}
	\label{fig:branching}
\end{figure}

Composition frameworks also allow users to declare branching behavior in a function composition. Branching behavior is achieved by declaring simple conditional expressions that assess a value received as input. We provide an example of branching behavior in Figure~\ref{fig:branching}. Unlike conditionals written as part of a normal function body, branching behavior defined in a platform function composition framework executes outside of the application, on resources owned and managed by the platform. To encode these conditionals in the event semantics, we create a virtual function whose COMPUTE step evaluates the conditional, then performs the event rule for the branch whose condition is met. For example, if the value of \texttt{doAuth} in an execution of the composition from Figure~\ref{fig:branching} were 1, the state transitions observed would be:

\small
\begin{align*}
& \mathbb{C}\mathbb{R}(f_{\mathit{step}}, x, \{\mathit{doAuth} : 1\}) & \qquad & \text{} \\
\implies & \mathbb{C}\mathbb{R}(f_{\mathit{step}}, x, v) \mathbb{F}(f_{\mathit{virt}}, \sigma[\mathit{doAuth} : 1], y) & \qquad &  \text{START($f_{\mathit{virt}}$)} \\
\implies & \mathbb{C}\mathbb{R}(f_{\mathit{step}}, x, v) \mathbb{F}(f_{\mathit{virt}}, \sigma' = \sigma[\mathit{cond} \leftarrow \sigma[\mathit{doAuth}] == 1], y) & \qquad &  \text{COMPUTE} \\
\implies & \mathbb{C}\mathbb{R}(f_{\mathit{step}}, x, v) \mathbb{F}(f_{\mathit{virt}}, \sigma', y) \mathbb{F}(f_{\mathit{auth}}, \sigma', y) & \qquad &  \text{INVOKE($f_{\mathit{auth}}$, v)} \\
\end{align*}
\normalsize

Since COMPUTE steps represent abstract local operations on a serverless function, they are sufficiently general to capture the evaluation of conditional logic. If the value of \texttt{doAuth} were 0, the series of transitions observed would be the same, though the function initialized by the final application of INVOKE would be $f_{\mathit{fail}}$.

%

\section{Container Prewarming}  \label{prewarming.sec}
In addition to their standard applications in program analysis, such information flow  analysis and  dead code detection, service call graphs have applications specific to serverless platforms, for example, container prewarming. Container prewarming is a strategy for mitigating delays associated with new container deployment during cold starts. Cold starts for JavaScript have previously been measured to incur as much as 644 ms of delay on AWS Lambda, and 9822 ms of delay on Azure~\cite{manner2018cold}. In prewarming, containers are initialized before they are needed and are kept warm by sending mock requests that trigger invocations at regular intervals. This strategy can be effective when the workload is predictable, where the correct number of containers can be kept warm. When workloads are intermittent or bursty, it may not be possible to predict the number of containers that are needed. Thus, this pre-warming approach can lead to wasted function invocations on unused containers or cold start latency penalties when too few containers are provisioned~\cite{ping2019mitigating}.

We propose an event-triggered prewarming approach that leverages our service call graphs. For each entry point to the application, e.g., each possible web request, the call graph identifies the chain of functions that may be triggered by that request. In the example in Figure~\ref{fig:galleria-comparison}, a web request triggers the \texttt{uploader} function, which leaders to a sequence of function executions, \texttt{rotator}, \texttt{resizer}, and \texttt{compressor}, each one triggering the next. In our scheme, as soon as \texttt{uploader} function is invoked, we send mock requests for the remaining three functions. If no containers are available for these functions, the requests will start the initialization of new containers , thus reducing any  cold start penalties.

\begin{table*}[]
	\caption{Startup Times for Serverless Functions in Galleria (in seconds)}
	\centering
	\begin{tabular}{@{}|l|r|r|r|r|r|r|@{}}
		\hline
		& \multicolumn{3}{|c|}{Mean}  &           \multicolumn{3}{|c|}{Median} \\
		\hline
		Function Name & Cold & Warm & Penalty & Cold & Warm & Penalty \\
		\hline
		uploader      & 2.562     & 2.122      &   0.440     &  2.562     & 2.125 & 0.437 \\ 
		\hline
		rotater       & 1.731     & 1.065      &   0.665     &  1.608     & 1.005 & 0.603 \\ 
		\hline
		resizer       & 1.425     & 1.329      &   0.095     &  1.515     & 1.079 & 0.436 \\ 
		\hline
		compressor    & 2.173     & 1.294      &   0.879     &  2.021     & 1.095 & 0.926 \\ 
		\hline
	\end{tabular} 
	\label{tab:coldstart}
\end{table*}

To demonstrate the effectiveness of event-triggered prewarming, we measure cold start and warm start times for the \texttt{galleria} application in Figure~\ref{fig:galleria-comparison} using CloudWatch logging.  We run \texttt{galleria} on AWS Lambda using 1536 MB containers for each serverless function. 
Container start times were calculated by measuring time elapsed from completion of the previous serverless function in the function chain to the start of the next function. For \texttt{uploader}, start time is the time elapsed since the web request was issued. Cold starts were triggered by leaving the application dormant for 50 minutes prior to the request. Each cold start was followed by five warm starts, each a single request spaced two minutes apart. 90 measurements were collected in total, 15 cold starts and 75 warm starts.
The measurements are shown in Table~\ref{tab:coldstart}.

In our experiments, the cold start penalty represents up to 45\% of total function execution time. Given a prewarming scheme that begins warming all functions in a chain upon a cold start of the first function, we calculate that end-to-end median cold start time is reduced from 7.706 seconds to 5.741 seconds, improving performance by nearly 25\%.

\end{document}